\documentclass[11pt]{article}

\usepackage{SIGACTNews}

\newcommand{\COLUMNsplit}{\vskip 2em\hrule \vskip 3em}

\newcommand{\COLUMNguest}[2]{%
  \begin{center}%
   {\LARGE\bf #1 \par}%
    \vskip 1.5em%
    {\Large
      \lineskip .75em%
      \begin{tabular}[t]{c}%
        #2
      \end{tabular}\par}%
    \vskip 1.5em%
  \end{center}%
  \par}

\usepackage{amsmath}
\usepackage{amsthm}

\usepackage{xspace}
\usepackage{graphicx}

\usepackage{amssymb}
\usepackage{amsfonts}
\usepackage{proof}
\usepackage{epsfig}
\usepackage{verbatim}

\newcommand{\LL}{\mathcal{L}}

\newcommand{\MM}{\mathcal{M}}

\renewcommand{\models}{\vDash}

\newcommand{\lam}{\lambda}

\newcommand{\aprolog}{{\ensuremath \alpha}{Pro\-log}\xspace}
\newcommand{\lprolog}{{\ensuremath \lam}{Pro\-log}\xspace}

\newcommand{\ab}[1]{\langle #1 \rangle}

\newcommand{\andd}{\wedge}
\newcommand{\orr}{\vee}
\newcommand{\impp}{\supset}
\newcommand{\ent}{\mathrel{{:}{-}}}

\newcommand{\nott}{\neg}
\newcommand{\true}{\top}
\newcommand{\false}{\bot}
\newcommand{\eq}{\approx}

\newcommand{\hyp}[2][]{\infer[#1]{#2}{}}

\newcommand{\new}[1][]{\reflectbox{\sf{#1}N}}

\newcommand{\fresh}{\mathrel{\#}}

\newcommand{\abs}[2]{{\ab{#1}{#2}}}

\newcommand{\name}[1]{\mathsf{#1}}
\newcommand{\Aa}{\name{a}}
\newcommand{\Ab}{\name{b}}
\newcommand{\Ac}{\name{c}}

\newcommand{\Ax}{\name{x}}
\newcommand{\Ay}{\name{y}}

\newcommand{\BA}{\mathbb{A}}

\newcommand{\tran}[2]{(#1~#2)}
\newcommand{\swap}[3]{(#1~#2)\act#3}

\newcommand{\act}{\boldsymbol{\cdot}}

\newcommand{\labelEq}[1]{\label{eqn:#1}}
\newcommand{\refEq}[1]{(\ref{eqn:#1})}

\newcommand{\labelFig}[1]{\label{fig:#1}}
\newcommand{\refFig}[1]{Figure~\ref{fig:#1}}

\newcommand{\labelProp}[1]{\label{prop:#1}}
\newcommand{\refProp}[1]{Proposition~\ref{prop:#1}}

\newcommand{\subs}[2]{\{#1/#2\}}
\newcommand{\exch}[2]{(#1{\leftrightarrow}#2)}

\newcommand{\seq}{\Rightarrow}
\newcommand{\nd}{\vdash}

\newcommand{\NPTIME}{\mathbf{NP}}

\newcommand{\FOLDNabla}{\ensuremath{FO\lambda^{\Delta\nabla}}\xspace}
\newcommand{\FOLNabla}{\ensuremath{FO\lambda^\nabla}\xspace}

\newtheoremstyle{example}{\topsep}{\topsep}%
     {}%
     {}%
     {\bfseries}%
     {.}%
     {10pt}%
     {\thmname{#1}\thmnumber{ #2}\thmnote{ (#3)}}%

\newtheorem{theorem}{Theorem}

\newtheorem{proposition}[theorem]{Proposition}

\theoremstyle{example}

\renewcommand{\phi}{\varphi}

\renewcommand{\bar}[1]{\overline{#1}}
\renewcommand{\vec}{\bar}
\newcommand{\ety}{\mathsf{exp}}
\newcommand{\vty}{\mathsf{var}}
\renewcommand{\subs}[2]{\{#2/#1\}}

\title{SIGACT News Logic Column 14}
\author{Riccardo Pucella\\
Northeastern University\\
Boston, MA 02115 USA\\
riccardo@ccs.neu.edu}
\date{}

\begin{document}

\SIGACTmaketitle

For this issue, James Cheney describes nominal logic, an approach to solve the
problems involved with reasoning about bindings in formal languages
that has been gaining in popularity in recent years, and surveys its
major application areas.

I am always looking for contributions. If you have any suggestion
concerning the content of the Logic Column, or---even better---if you
would like to contribute by writing a column, please feel free to get
in touch with me. (Please note that my contact information has
changed.)

\COLUMNsplit

\COLUMNguest{Nominal Logic and Abstract Syntax\footnote{\copyright{} James Cheney, 2005.}}
   {James Cheney\\University of Edinburgh}

\section{Introduction}

A great deal of research in programming languages, type theory, and
security is based on proving properties such as strong normalization,
type soundness or noninterference by induction or co-induction on the
structure of typing derivations, operational semantics rules, or other
syntactic constructs.  Such proofs are essentially combinatorial in
nature, usually involving $O(n^p)$ cases, where $n$ is the number of
syntactic constructs, typing rules, operational transitions, etc., and
$p$ is small.  Usually, only a small number of cases are
``interesting'', and published proofs often give only a few
illustrative cases.  This provides little assurance that a proof is
correct.  It is widely felt that machine assistance for constructing
such proofs is
desirable~\cite{Pfenning:HandbookAR:framework:2001,poplmark05tphol}.
Providing such assistance is severely complicated by the problem of
dealing with names and binding in abstract syntax.

Logicians since Frege have grappled with the problem of dealing with
the syntax of logical expressions.  In mathematical logic textbooks it
is not unusual to see a formal definition of the \emph{concrete
  syntax} of the object-language as a particular set of strings over
an alphabet including not only variables, function and predicate
symbols, and logical connectives, but also punctuation such as
parentheses, brackets, and commas.  Then various technical lemmas such
as the fact that a string may represent at most one formula, that
parentheses match, etc. are proved, along with structural induction
principles.  These results are necessary to show that the use of
strings and punctuation to represent formulas satisfies our intuitions
about the ``real'' structure of formulas.

However, we now have both a highly developed theory and advanced
programming techniques for taking care of these syntactic details
automatically; many high-level programming languages (such as Prolog
and ML) provide advanced features for parsing strings into abstract
syntax trees and computing with the results.  Nowadays it is more
common (and by far more agreeable) to specify a logical or
mathematical language using \emph{abstract syntax}, that is, as a set
of abstract syntax trees defined by an inductive construction.  This
has the positive effect of isolating the low-level technical details
of parsing from the high-level hierarchical representations of terms
which are most convenient for reasoning.  Moreover, the theory of
abstract syntax trees, term languages, etc., is now well-understood, so
that a large number of definitions and results are standard and can be
reused for any term language (and are usually taken for granted).

While this approach works well for languages with variables,
constants, and operator symbols, it does not work so well as soon as
variable binding enters the picture.  This is because variable binding
and substitution interact in complex ways.  Subtle errors can arise if
care is not taken with definitions; this problem has plagued both
famous logicians\footnote{including Hilbert and
  Ackermann~\cite{hilbert28} among others, according to
  Stoy~\cite{stoy81}} and well-known programming
languages.\footnote{for example, LISP's well-known ``dynamic variable
  scoping'' bug~\cite{barendregt97bsl}}

It is standard practice in mathematical logic to assume that there is
some infinite set $V$ of variable symbols, for example integers or
strings, and to treat binding term constructors as ordinary function
symbols taking variables as arguments or parameters.  For example, the
$\forall$ symbol in a universally quantified formula $\forall x.\phi$
may be viewed, from a syntactic point of view, as a binary function
symbol $forall : V \times Prop \to Prop$, or as a family of unary
formula constructors $(\forall_x : Prop \to Prop \mid x \in V)$ taking
a formula as an argument.  If one of these approaches is employed,
then a number of basic syntactic definitions and results (such as the
``alphabetic variance'' or $\alpha$-equivalence relation,
capture-avoiding substitution function and related lemmas) again need
to be proved in order to establish that this approach to implementing
binding matches our intuitive understanding.

Once these low-level details of binding have been presented and proved
correct, mathematical rigor is usually reserved for high-level issues,
and low-level syntactic and binding issues are left implicit.  For
example, the \emph{Barendregt Variable Convention} is often taken for
granted in mathematical exposition.  After defining capture-avoiding
substitution, renaming, and $\alpha$-equivalence and proving their
properties in detail, Barendregt states:
\begin{quote} 2.1.13.
  \textsc{Variable Convention}. If $M_1,\ldots,M_n$ occur in a certain
  mathematical context (e.g., definition, proof), then in these terms
  all bound variables are chosen to be different from free
  variables.~\cite{barendregt84}
\end{quote}
At best, paper definitions or proofs signal the use of such a
convention with a ``without loss of generality''; at worst, the
convention (and the argument that its use is sound) is implicit.
While clear enough for human readers, however, such conventions still
leave a considerable gap between mathematical exposition and correct
formalizations or computer implementations of programming languages
and logics involving names and binding~\cite{vaninwegen96phd,vestergaard03ic}.

In this column, I will first survey the state of the art for solving
these problems, and then present a new approach called \emph{nominal
  abstract syntax} that has gained popularity since its introduction
six years ago by Gabbay and Pitts.  I will then discuss applications
and future directions for this work.

\section{Approaches to dealing with names and binding}

\subsection{First-order abstract syntax}

In programming, abstract syntax with binding is usually implemented
using ordinary abstract syntax techniques, and then defining
capture-avoiding renaming/substitution and $\alpha$-equivalence
explicitly; that is, by making all the implicit syntactic
manipulations explicit.  This \emph{first-order abstract syntax}
approach requires explicit management of fresh name generation (e.g.
using a side-effecting $gensym$ function) as well as writing a lot of
repetitive ``boilerplate'' code.  Despite these drawbacks, it is by
far the most popular technique for real compilers, interpreters,
theorem provers, and other symbolic programs.  However, reasoning
about languages defined in this way is considered by most experts in
this area to be impractical for all but the simplest examples, because
of the large number of intermediate reasoning steps, renaming and
substitution lemmas, etc.  that must be verified.

\subsection{Name-free approaches}
Another popular technique for managing abstract syntax with binding is
to use a \emph{name-free} notation for functions.  Name-free
approaches have a long history, beginning with Sch\"onfinkel's
development of combinatory logic in the
1920s~\cite{schoenfinkel67fftg}, and have had considerable influence
on both theory and practice of logic and programming.
Sch\"onfinkel~\cite{schoenfinkel67fftg} and later Curry and
Feys~\cite{curry30ajm,curry58} developed \emph{combinatory logic}, a
logic of applicative expressions defined using rewriting rules.  In
combinatory logic, a $\lambda$-term such as $\lambda x. \lambda y. x
y$ can be expressed as the combinator expression
$S(S(KS)(S(KK)I))(KI)$; here $S$, $K$, and $I$ are basic functional
expressions with the same meaning as $\lambda xyz. (xz)(yz)$, $\lambda
xy.x$, and $\lambda x.x$, respectively. N. G.  de Bruijn~\cite{debruijn72im} proposed two
encodings (often called \emph{de Bruijn indices} and \emph{de Bruijn
  levels}) for the $\lam$-calculus which neatly circumvent the
difficulties arising from $\alpha$-equivalence by representing
variables as integer references (or \emph{pointers}) to their binding
sites.  For example, the de Bruijn index version of both $\lambda x.
\lambda y. x y$ and $\lambda z. \lambda w. z w$ is $\lambda \lambda 2
1$. Thus, $\alpha$-equivalence collapses to syntactic equality.
\emph{Stoy diagrams}~\cite{stoy81} are a graphical representation of
$\lambda$-terms often used to explain variable binding; an example is
shown in \refFig{stoy}.

Combinators and de Bruijn index representations are powerful and
useful ideas; the former serves as the basis for efficient functional
language implementations~\cite{cousineau87scp}, while the latter has
been used as an efficient internal representation in many theorem
provers (beginning with de Bruijn's AUTOMATH) and in efficient
functional programming implementation techniques such as
\emph{explicit substitutions}~\cite{abadi91jfp}.  Nevertheless,
combinators can increase the size of an expression exponentially, and
neither encoding is human-readable so neither is well suited for
high-level programming or reasoning tasks.  Moreover, name-free
approaches by definition do not provide any assistance for dealing
with free names.

\begin{figure}
\begin{center}
\includegraphics{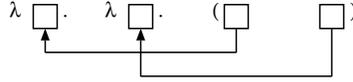}
\end{center}
\caption{Stoy diagram for $\lambda x. \lambda y. (x~y)$.}\labelFig{stoy}
\end{figure}

\subsection{Higher-order abstract syntax}
An elegant alternative is to employ \emph{higher-order abstract
  syntax} (HOAS)~\cite{pfenning89pldi}.  In this technique, we work
with an enriched meta-language that provides some form of binding
(such as the typed lambda-calculus).  Variables and binding term
constructors at the object level are encoded as variables and
higher-order constants in the metalanguage.  This is best illustrated
by an example.  Using higher-order abstract syntax, a quantified
formula like $\forall x{:}\mathbb{N}.P(x)$ would be encoded as
$forall~(\lambda x.  P~x)$, where $forall : (Nat \to Prop) \to Prop$,
$P : Nat \to Prop$. This powerful idea, first used in Church's
higher-order logic~\cite{church40jsl}, is used in many advanced
programming languages and logical frameworks (e.g.
\lprolog~\cite{nadathur98higher} and Twelf~\cite{pfenning99system}
among others~\cite{barzilay02tphol,despeyroux95tlca}).

However, in my opinion, higher-order abstract syntax is not without
drawbacks, both from the point of view of \emph{programming with} and
\emph{reasoning about} languages involving binding.  The problems can
be broken down into five areas:
\begin{enumerate}
\item Higher-order abstract syntax is based on complex semantic and
  algorithmic foundations (higher-order logic~\cite{church40jsl},
  recursive domain equations~\cite{smyth82siamjc}, higher-order
  unification~\cite{huet75tcs}) so requires a fair amount of ingenuity
  to learn, implement and
  analyze~\cite{fiore99lics,hofmann99lics,schurmann01tcs}.
\item Properties of the metalanguage (such as weakening and
  substitution lemmas) are inherited by object languages, whether or
  not this is desirable; this necessitates modifications to handle
  logics and languages with unusual behavior. Examples include linear
  logic~\cite{cervesato02ic} and concurrency~\cite{watkins03types}.
\item Variable names are ``second class citizens''; they only
  represent other object-language expressions and have no existence or
  meaning outside of their scope.  This complicates formalizing
  languages with generativity (for example, datatype names in ML),
  program logics with mutable variables such as Hoare
  logic~\cite{mason87lfcs} or dynamic logic~\cite{honsell96types} and
  translations such as closure conversion that rely on the ability to
  test for name-equality.
\item Higher-order language encodings are often so different from
  their informal ``paper'' presentations that proving ``adequacy''
  (that is, equivalence of the encoding and the real language) is
  nontrivial, and elegant-looking encodings can be incorrect for subtle
  reasons. Hannan~\cite{hannan95wta} developed and proved partial
  correctness of a closure conversion translation in LF, but did not
  prove adequacy of the encoding; careful inspection suggests that it
  is not adequate.  Abel~\cite{abel01merlin} investigated an elegant
  and natural-seeming but inadequate third-order and less elegant but
  adequate second-order HOAS encoding of the $\lambda\mu$-calculus.
\item Higher-order abstract syntax is less expressive than first-order
  abstract syntax: it apparently cannot deal with situations involving
  ``open terms'' mentioning an indefinite number of free variables.
  For example, HOAS apparently cannot model the behavior of ML-style
  let-bound polymorphism as usually implemented~\cite{harper90tr},
  though a simulation is possible~\cite{hannan88meta}.
\end{enumerate}
To be fair, \emph{when it works}, higher-order abstract syntax is
highly satisfying and clearly superior to first-order abstract syntax,
and research on higher-order abstract syntax has shown that many of
these problems can be alleviated.  In addition, any of these properties
can also be seen in a more positive light:
\begin{enumerate}
\item[I.] Higher-order abstract syntax is based on powerful and elegant
  semantic and algorithmic foundations (higher-order logic, recursive
  domain equations, higher-order unification) involving deep ideas of
  computer science and logic.
\item[II.] Properties of the metalanguage (such as weakening and
  substitution lemmas)  are inherited by object languages, 
  thus saving a lot of work re-proving them for typical languages.
\item[III.] Programmers do not have to painstakingly re-implement
  efficient fresh name generation, $\alpha$-equivalence, or
  capture-avoiding substitution operations, freeing them to focus on
  high-level problems.
\item[IV.] Higher-order language encodings encourage ``refactoring''
  object-languages in a way that makes all variable binding explicit;
  the results are often much more elegant and uniform than ``paper''
  versions.
\item[V.] Higher-order abstract syntax encourages ``one binding at a
  time'' definitions and proofs, and discourages complicated (and
  frequently unnecessary) reasoning about open terms.
\end{enumerate}
I am not arguing that points (1--5) are right and (I--V) are wrong (or
vice versa); both views have merit.  Anyone interested in formalizing
programming languages or logics should consider higher-order abstract
syntax, because it enjoys several mature implementations such as Twelf
and \lprolog.  Nevertheless, I believe it is worthwhile to investigate
alternatives.

\section{Nominal abstract syntax}

Recently, Gabbay and Pitts~\cite{gabbay99lics,gabbay02fac} have
developed an alternative approach to abstract syntax with binding.
This approach is based on the idea of taking names to be an abstract
but first-class data type, and name-binding (or name-abstraction) to
be an abstract data type construction involving the type of names and
an arbitrary type.  Thus, as in first-order abstract syntax, names
denote semantic values, and are not just syntactic entities; on the
other hand, like higher-order abstract syntax, access to the internal
representations of names and name-binding operations is restricted so
that low-level implementation issues are separated from high-level
concerns.  I call this approach \emph{nominal abstract syntax},
because of its focus on names.

The key technical insight in nominal abstract syntax is the fact that
one-to-one (or, equivalently, invertible) renamings can play a central
role in explicating name-binding (and, in fact, many other uses of
names).  At an intuitive level, the reason is that in many situations,
the only properties of names that are of interest are
\emph{equality/inequality} among names and \emph{freshness}, that is,
the property that a name does not appear ``free'' in a term.
Non-invertible capture-avoiding renamings do preserve equality (e.g.,
$t = u$ implies $t[x/y] = u[x/y]$) but may \emph{not} preserve
inequality or freshness.  For example, $x \neq y$ but $x[x/y] = x =
y[x/y]$, and $x \not\in FV(\lambda x.f~x~y)$ but $x[x/y] = x \in
FV(\lambda x'.f ~x'~x) = FV((\lambda x.f~x~y)[x/y])$.  Invertible
renamings, on the other hand, preserve all of these properties: for
example, writing $(-)\exch{x}{y}$ for simultaneous capture-avoiding
substitution of $x$ for $y$ and $y$ for $x$, we have $x\exch{x}{y} = y
\neq x = y\exch{x}{y}$ and $x\exch{x}{y} = y \not\in FV(\lambda x'.
f~x'~x) = FV((\lambda x.f~x~y)\exch{x}{y})$.

Although an approach based on invertible renamings may seem unnatural to modern
logicians and computer scientists, Gabbay and Pitts were not the first
to recognize the importance of invertible renamings.  Permutations
are frequently used to define the equivalence class of
\emph{alphabetic variants} of an object (e.g., in logic programming,
two logic program clauses or unifiers which differ only by a
permutation of variables are equivalent; graphs and automata are
considered equivalent up to invertible renaming of state names).
Also, in a formalization of the $\lambda$-calculus, McKinna and
Pollack~\cite{mckinna99jar} identified invertible renamings as an
important concept that can be used to define $\alpha$-equivalence.

In fact, the basic idea of using \emph{one-to-one renamings} to
understand name-binding dates to Frege's \emph{Be\-griffs\-schrift},
the first systematic treatment of  symbolic predicate logic. In
Frege's work, bound names were syntactically distinguished from
unbound names: the former were written using German letters
$\mathfrak{a,b,c}$, and the latter using Roman (italic) letters
$x,y,z$.  For Frege, bound names were subject to renaming using the
following principle:
\begin{quote}
  $\ldots$ Replacing a German letter everywhere in its scope by some
  other one is, of course, permitted, so long as in places where
  different letters initially stood different ones also stand
  afterward.  This has no effect on the content.~\cite{frege67}
\end{quote}
On the other hand, Frege did not give a completely explicit formal
treatment of substitution, and as a result much of the complexity
resulting from the interaction of substitution and name-binding was
hidden.

Gabbay and Pitts' original approach was based on ideas from
Fraenkel-Mostowski \emph{permutation models} of set theory (FM set
theory)~\cite{fraenkel67ac,felgner70models,jech77about}, originally
developed as an early attempt to prove the independence of the Axiom
of Choice from ZF-set theory.  In fact, Gabbay and Pitts' work was
carried out in a form of FM set theory that does not satisfy the Axiom
of Choice.  While this is a very interesting approach, it has led many
observers to believe that nonstandard FM set theory and rejection of the
Axiom of Choice is \emph{necessary} for working with nominal abstract
syntax, not just \emph{sufficient}.

This is not the case; Pitts~\cite{pitts03ic} showed that the basic
principles of nominal abstract syntax can be formalized as
\emph{nominal logic}, an extension of typed first-order equational
logic that can be analyzed within ZFC just like any other logic.  In
this approach, there is no conflict with the Axiom of Choice or the
mainstream foundations of mathematics.  On the other hand, nominal
logic lacks some of the Choice-like properties of first-order logic
(such as unrestricted Skolemization), but this is not a foundational
problem.

In the rest of this section I will provide a brief overview of nominal
abstract syntax and nominal logic, followed by an example of reasoning
in nominal logic.  Much more detail can be found in the
papers~\cite{pitts03ic, urban04tcs,gabbay04lics,cheney05fossacs}.

\subsection{Nominal logic}

The key ingredients of nominal logic are:
\begin{itemize}
\item a syntactic class of \textbf{names} $\Aa,\Ab,\ldots \in \BA$,
  partitioned into name-types $\nu,\nu',\ldots$,
\item a \textbf{swapping} operation $\swap{a}{b}{t}$ that swaps
  two names of the same type within an value,
\item a \textbf{freshness} relation $a\fresh t$ that relates a
  name to a value when the name does not appear ``free'' in the value,
\item an \textbf{abstraction} operation $\abs{a}{t}$ that binds a
  name within an value, and admits equality up to
  $\alpha$-equivalence, and
\item a self-dual \textbf{new-quantifier} $\new a{:}\nu.\phi$ that
  quantifies over \emph{all fresh names} (equivalently, \emph{some
    fresh name}) of type $\nu$.
\end{itemize}
In addition, nominal logic embodies two key logical principles:
\begin{itemize}
\item {\bf Fresh name generation.}  A name fresh for any value
  (or for each of finitely many values) can always be found.
\item {\bf Equivariance.}  Relations are invariant up to swapping; the
  choice of particular names in a formula is irrelevant.
\end{itemize}

The syntax of nominal logic is as follows:
\[\begin{array}{lrcl}
\text{(Types)}&  \sigma &::=& \nu \mid \delta \mid \abs{\nu}{\sigma}\\
\text{(Contexts)}&  \Sigma &::=& \cdot \mid \Sigma,x{:}\sigma \mid \Sigma\#\Aa{:}\nu\\
\text{(Terms)}&  t &::=& \Aa \mid c \mid f(\vec{t}) \mid x \mid \swap{a}{b}{t} \mid \abs{a}{t}\\
\text{(Atomic formulas)}&  A &::=&  p(\vec{t}) \mid t \eq u \mid a \fresh t\\
\text{(Formulas)}&  \phi &::=& A \mid \phi \impp \psi \mid \bot \mid \forall x{:}\sigma.\phi \mid \new a{:}\nu.\phi
\end{array}\]
A \emph{language} $\LL$ consists of a set of \emph{data types} $\delta$,
\emph{name types} $\nu$, \emph{constants} $c:\delta$, \emph{function
  symbols} $f:\vec{\sigma} \to \delta$, and \emph{relation symbols}
$p:\vec{\sigma}\to o$.
Well-formed terms and atomic formulas are defined as follows:
\[\begin{array}{c}
\infer{\Sigma \nd \Aa:\nu}{\Aa:\nu \in\Sigma}\quad
\infer{\Sigma \nd x:\sigma}{x:\sigma\in\Sigma}\quad
\infer{\Sigma \nd c:\delta}{c:\delta\in\LL}\quad
\infer{\Sigma \nd f(\vec{t}):\delta}{f:\vec{\sigma}\to\delta\in\LL & \Sigma \nd \vec{t}:\vec{\sigma}}\quad
\infer{\Sigma \nd \abs{a}{t}:\abs{\nu}{\sigma}}{\Sigma\nd a : \nu & \Sigma \nd t : \sigma}\\
\infer{\Sigma \nd \swap{a}{b}{t}:{\sigma}}{\Sigma\nd a : \nu &\Sigma\nd b : \nu & \Sigma \nd t : \sigma}\quad
\infer{\Sigma \nd t \eq u: o}{\Sigma \nd t:\sigma & \Sigma \nd u:\sigma}\quad
\infer{\Sigma \nd a \fresh t: o}{\Sigma \nd a:\nu & \Sigma \nd t:\sigma}\quad
\infer{\Sigma \nd p(\vec{t}): o}{p: \vec{\sigma} \to o\in \LL &\Sigma \nd \vec{t}:\vec{\sigma}}
\end{array}\]
Here we write $o$ for the type of propositions; however,
quantification over types mentioning $o$ is not allowed.
Well-formedness for the first-order connectives are defined in the
usual way; well-formedness for $\new a{:}\nu.\phi$ is defined as for
$\forall a{:}\nu.\phi$.  Other formulas such as truth $\true$,
conjunction $\phi \andd \psi$, disjunction $\phi \orr \psi$, logical
equivalence $\phi \iff \psi$, and existential quantification $\exists
x{:}\sigma.\phi$ are defined as usual in classical logic.

While ordinary capture-avoiding renaming/substitution needs to be
defined carefully to prevent variable capture, invertible renamings
can be defined by a simple structural induction:
\[
\begin{array}{rcl}
  \swap{\Aa}{\Ab}{\Aa} &=& \Ab\\
  \swap{\Aa}{\Ab}{\Ab} &=& \Aa\\
  \swap{\Aa}{\Ab}{\Aa'} &=& \Aa' \quad(\Aa \neq \Aa' \neq \Ab)
\end{array}
\qquad 
\begin{array}{rcl}
  \swap{\Aa}{\Ab}{c} &=& c\\
  \swap{\Aa}{\Ab}{f(t_1,\ldots,t_n)} &=& f(\swap{\Aa}{\Ab}{t_1},\ldots,\swap{\Aa}{\Ab}{t_n})\\
  \swap{\Aa}{\Ab}{\abs{\Aa'}{t}} &=& \abs{\swap{\Aa}{\Ab}{\Aa'}}{\swap{\Aa}{\Ab}{t}}
\end{array}
\]
Note that the names appearing in ``binding'' positions of abstractions
are not $\alpha$-renamed prior to applying a swapping; instead, the
swapping is applied to both the name and body.  This would be
incorrect for a \emph{non-invertible} renaming, e.g. 
\[(\lambda x.
y)[x/y] = \lambda x'.x \neq \lambda x.x = \lambda (x[x/y]). y[x/y]
\]
however, invertibility
ensures that ``variable capture'' is avoided:
\[\swap{\Aa}{\Ab}{(\abs{\Aa}{\Ab})} = \abs{\Ab}{\Aa} =
\abs{\swap{\Aa}{\Ab}{\Aa}}{\swap{\Aa}{\Ab}{\Ab}}
\]
That is,
\emph{invertible renamings are inherently capture-avoiding}.

Next, we define what it means for a name to be independent of (or
\emph{fresh} for) a term.  Intuitively, a name $\Aa$ is fresh for a
term $t$ (that is, $\Aa \fresh t$) if $t$ possesses no occurrences of
$\Aa$ unenclosed by an abstraction of $\Aa$.  We define this using the
following inference rules:
\[\begin{array}{c}
\infer{\nd \Aa \fresh \Ab}{(\Aa \neq \Ab)}\quad
\infer{\nd\Aa \fresh c}{}\quad
\infer{\nd\Aa \fresh f(t_1,\ldots,t_n)}{\nd\Aa \fresh t_i & (i = 1,\ldots,n)}\quad
\infer{\nd\Aa \fresh \abs{\Ab}{t}}{\nd\Aa \fresh \Ab & \nd\Aa \fresh t}\quad
\infer{\nd\Aa \fresh \abs{\Aa}{t}}{}
\end{array}\]
We sometimes refer to the set of ``free'' names of a term $FN(t) =
\BA - \{\Aa \mid \Aa \fresh t\}$ as its \emph{support}.

Finally, we define an appropriate equality relation on nominal terms
that identifies abstractions up to ``safe'' renaming.
\[\begin{array}{c}
\hyp{\nd\Aa \eq \Aa}\quad
\hyp{\nd c \eq  c}\quad
\infer{\nd f(t_1,\ldots,t_n) \eq f(u_1,\ldots,u_n)}{\nd t_i \eq u_i & (i = 1,\ldots,n)}\quad
\infer{\nd\abs{\Aa}{t} \eq \abs{\Aa}{u}}{\nd t \eq u }\quad
\infer{\nd\abs{\Aa}{t} \eq \abs{\Ab}{u}}{\nd\Aa \fresh u & \nd t \eq \swap{\Aa}{\Ab}{u}}
\end{array}\]
For example, $\abs{\Aa}f(\Aa,\Ab) \eq \abs{\Ac}{f(\Ac,\Ab)} \not\eq
\abs{\Ab}{f(\Ab,\Aa)}$; the first equation is derived as follows:
\[\infer{\nd \abs{\Aa}f(\Aa,\Ab) \eq \abs{\Ac}{f(\Ac,\Ab)}}
{\infer{\nd \Aa \fresh f(\Ac,\Ab)}{\infer{\nd \Aa\fresh\Ac}{} 
    & \infer{\nd \Aa\fresh\Ab}{}} 
  & \infer{\nd f(\Aa,\Ab) \eq {f(\Aa,\Ab)}}
  {\infer{\nd \Aa\eq\Aa}{} 
    & \infer{\nd \Ab\eq\Ab}{}}}
\]
The second rule for abstraction may seem asymmetric because we do not
check that $\Ab \fresh t$. In fact, this check is redundant: If $\Aa
\fresh t$ and $t \eq \swap{\Aa}{\Ab}{u}$, then $\Aa \fresh
\swap{\Aa}{\Ab}{u}$; by applying the swapping $\tran{\Aa}{\Ab}$ to
both sides, we get $\Ab \fresh u$, since $\swap{\Aa}{\Ab}{\Aa} = \Ab$
and $\swap{\Aa}{\Ab}{\swap{\Aa}{\Ab}{u}} =u$.  (It is
straightforward to show that $a \fresh t$ implies $\swap{b}{b'}{a}
\fresh \swap{b}{b'}{t}$ for any $a,b,b',t$).

Nominal logic proper consists of first-order logic extended with a
first-order axiomatization of swapping, freshness, and abstraction,
and with a new form of quantified formula, $\new a.\phi$.  For the
purposes of this column, we restrict attention to \emph{term models}
of nominal logic, in which function symbols and constants are
interpreted as themselves (with swapping and abstraction having the
special meanings given by the above rules).  The intended objects of
study in nominal logic are usually terms, and focusing on term models
means that we can avoid some model-theoretic subtleties (for more on
this issue, see~\cite{cheney05jsl}).

Since there is no choice in the interpretation of the constants and
function symbols, a term model $\MM$ can be represented as a set of
ground atomic formulas $A$; we require that this set be closed under
renaming, so that $\swap{a}{b}{A} \in \MM$ if and only if $A \in \MM$.  For term
models, the semantics of closed nominal logic formulas can be defined
as follows:
\[\begin{array}{lcl}
 \MM\not\models \false\\
 \MM \models A &\iff& A \in \MM\\
\MM\models t \eq u &\iff& \nd t \eq u\\
 \MM\models a \fresh u &\iff& \nd a \fresh u\\
 \MM\models \phi\impp \psi &\iff& \text{$\MM\models \phi$  implies  $\MM\models \psi$}\\
 \MM\models \forall x{:}\sigma.\phi(x) &\iff& \text{$\MM\models \phi(t)$ for every $t:\sigma$}\\
 \MM\models \new a{:}\nu.\phi(a) &\iff& \text{$\MM\models \phi(\Aa)$ for some $\Aa :\nu \not\in FN(\phi(a))$}
\end{array}\]
As an example, we consider the following (valid) property of nominal logic:
\[\models \forall a, b{:}\nu, x{:}\sigma.~ a \fresh x \andd b \fresh x \impp \swap{a}{b}{x} \eq x\]
To verify the validity of this formula, it suffices to show by
structural induction on terms $t$ that for any concrete names
$\Aa,\Ab\fresh t$, $\swap{\Aa}{\Ab}{t} \eq t$.  This is
straightforward for all but the abstraction cases; for abstractions,
if the abstracted name is $\Aa$ or $\Ab$, then the second equality
rule for abstractions must be used.

We next establish further properties of nominal logic mentioned above.
In particular, it is easy to show by induction (on $n$-tuples of terms
$\vec{t}$) that the freshness principle
\[(F)\qquad\forall \vec{x}.~\exists a{:}\nu. a \fresh \vec{x}
\]
is valid.  Similarly, it is straightforward to show by induction on $\phi$ 
that the equivariance principle
\[ (EV)\qquad \phi \iff \swap{a}{b}{\phi}
\]
is valid.  We also made the (possibly counterintuitive) claim that
$\new$ is self-dual; that is, $\nott \new a.\phi(a) \iff \new a.\nott
\phi(a)$.  To prove this, suppose $\MM \not\models \new a.\phi(a)$.
Then for \emph{every} $\Aa \not\in FN(\phi(a))$, $\MM \not\models
\phi(\Aa)$.  Since $FN(\phi(a))$ is finite and $\BA$ is infinite, we
may choose a particular $\Aa\not\in FN(\phi(a))$ such that $\MM
\not\models \phi(\Aa)$.  Hence, $\MM \models \nott \phi(\Aa)$, and since
$\Aa\not\in FN(\phi(a))$ we can conclude that $\MM \models \new a.\nott
\phi(a)$.  Using similar reasoning it is not difficult to show that
\[\new a.\phi(a,\vec{x}) \iff \exists a.a \fresh \vec{x} \andd \phi(a,\vec{x}) \iff \forall a. a \fresh \vec{x} \impp \phi(a,\vec{x})\]

Because of this self-duality property, we can use particularly simple 
proof rules for the $\new$-quantifier.  For example, sequent-style
rules have the following form:  
\[\infer[\new R]{\Sigma:\Gamma \seq \new a.\phi(a)}{\Sigma\#\Aa:\Gamma \seq \phi(\Aa)} 
\qquad 
\infer[\new L]{\Sigma:\Gamma ,\new a.\phi(a)\seq  C}{\Sigma\#\Aa:\Gamma,\phi(\Aa) \seq C }\]
Here, the context $\Sigma$ contains variables (introduced as $\forall$
or $\exists$ parameters) and names (introduced by the $\new$-rules).
The context $\Sigma\#\Aa$ indicates that $\Aa$ is assumed to be
distinct from all names in $\Sigma$ and fresh for all values of
variables in $\Sigma$.  intuitively, these rules state that to either
prove a $\new$-quantified conclusion or make use of a
$\new$-quantified hypothesis, it suffices to instantiate the
conclusion or hypothesis with a completely fresh name and proceed.  In
the example to follow, we won't be completely formal about proofs in
nominal logic; instead we will reason at the semantic level about term
models.  More detail about the proof theory can be found in the 
papers~\cite{gabbay04lics,cheney05fossacs}.

\subsection{A theory of the syntax of the $\lambda$-calculus}

As an example of the expressiveness of nominal logic, we show how the
abstract syntax of the $\lambda$-calculus~\cite{barendregt84} can be formalized as a theory
$\Gamma_\Lambda$ of nominal logic.  We also prove some simple
properties concerning capture-avoiding substitution.  We consider a
language including one data type $\ety$ for $\lambda$-terms, one name-type
$\vty$ for $\lambda$-calculus variable names, and the following
function symbols:
\begin{eqnarray*}
var &:& \vty \to \ety \\ lam &:& \abs{\vty}{\ety}\to \ety \\ app &:& \ety \times \ety \to \ety
\end{eqnarray*}
We use $a,b,c$ for variables of type $\vty$, and $M,N$ for variables
of type $\ety$.  Since we are interpreting nominal logic over
syntactic models only, we assume the following axioms expressing that
$var$, $lam$, and $app$ are injective functions, and their ranges are
disjoint:
\begin{gather}
var(a) \eq var(b) \impp a \eq  b\labelEq{var-inj}\\
app(M,N) \eq  app(M',N') \impp M \eq M' \andd N \eq N'\labelEq{app-inj}\\
lam(M) \eq  lam(M') \impp M \eq  M'\labelEq{lam-inj}\\
var(a) \not\eq app(M,N)\labelEq{var-app-neq}\\
var(a) \not\eq lam(M)\labelEq{var-lam-neq}\\
app(M,M') \not\eq lam(N)\labelEq{app-lam-neq}
\end{gather}

Let $P(x)$ be a formula with a free parameter $x{:}\ety$ (and
possibly other parameters). We can express a structural induction
principle over expressions as follows:
\[
\begin{array}{lcl}
(\Lambda_{ind})&&(\forall a{:}\vty.~P(var(a)))\\
&\andd&(\forall M,N{:}\ety.~ P(M) \andd P(N) \impp P(app(M,N)))\\
 &\andd&(\new a{:}\vty. \forall M{:}\ety.~ P(M) \impp P(lam(\abs{a}{M})))\\
&\impp& \forall x{:}\ety.~ P(x)
\end{array}
\]
Note that $FV(\forall M{:}\ety.~ P(M) \impp P(lam(\abs{a}{M}))) = \{a\}
\cup FV(P)$, so the $\new$-quantified name $a$ in the third case must
be fresh for any additional parameters of $P$.

The axioms \refEq{var-inj}--\refEq{app-lam-neq} together with all 
instances of $(\Lambda_{ind})$ form the theory $\Gamma_\Lambda$.  This
theory is valid in any term model $\MM$ for the language $\Lambda$.  
Moreover, we can extend the language with additional atomic formulas
defined by Horn clauses, as a consequence of a version of Herbrand's 
Theorem  for nominal logic~\cite{cheney05jsl}:
\begin{theorem}
Let $R_1,\ldots,R_n$ be fresh relation symbols and let $\Gamma$ be
a set of nominal Horn clauses, that is, closed formulas of the form
\[\new \vec{a}.\forall \vec{x}. A_1(\vec{a},\vec{x})\andd\ldots\andd A_n(\vec{a},\vec{x}) \impp
R_i(\vec{t}(\vec{a},\vec{x}))\]
where $A_1,\ldots,A_n$ are either freshness, equality, or $R_i$
formulas.  Then $\Gamma$ has a unique least term model $\MM$.
\end{theorem}
Nominal Horn clauses are also written in a Prolog-like notation, in
which $\new$-quantified variables are replaced by constant names:
\[R_i(\vec{t}(\vec{\Aa},\vec{x})) \ent A_1(\vec{\Aa},\vec{x}),\ldots, A_n(\vec{\Aa},\vec{x})\]
For example, \refFig{aprolog} shows a typechecking judgment and a
relation defining capture-avoiding substitution.  (We also use
Prolog-like notation for lists and a list membership predicate $mem$.)
Reading $b \fresh N$ as $b \not\in FV(N)$, the axioms for $subst$
correspond precisely to Barendregt's relational definition of
capture-avoiding substitution~\cite{barendregt84}.

\begin{figure}[tb]
\[\begin{array}{lclcl}
  typ(G,var(X), T)          &\ent& mem((X,T),G).\\
  typ(G,app(M,N),T')    &\ent& typ(G,M,arr(T,T')), typ(G,N,T).\\
  typ(G,lam(\abs{\Aa}M),arr(T,T')) &\ent& \Ax \fresh G,
  typ([(\Aa,T)|G],M,T').\\\\
  subst(var(\Aa),P,\Aa,P).\\
  subst(var(\Ab),P,\Aa,var(\Ab)).\\
  subst(app(M,N),P,\Aa,app(M',N')) &\ent& subst(M,P,\Aa,M'), subst(N,P,\Aa,N').\\
  subst(lam(\abs{\Ab}M),P,\Aa,lam(\abs{\Ab}M')) &\ent& \Ab \fresh P,subst(M,P,\Aa,M').
\end{array}\]
\caption{Simple Horn clause definitions}\labelFig{aprolog}
\end{figure}

It is inconvenient to work exclusively with relations, so we introduce 
a recursive definition principle which justifies adding function symbols
to the language.  First, we observe that nominal logic has a limited 
Skolemization property:
\begin{theorem}
  If $\MM \models \forall x{:}\sigma. \exists! y{:}\sigma'. F(\vec{x},y)$, then
  we may consistently extend the language with a constant $f:\sigma
  \to \sigma'$ satisfying $\forall x{:}\sigma.F(x,f(x))$.
\end{theorem}
Suitable generalizations to multiple-argument functions also hold.

It is not difficult to show that the $subst$ relation defined in
\refFig{aprolog} is total and functional in its first three arguments,
so we can Skolemize it as $-\subs{-}{-} : \ety \times \ety \times \vty
\to \ety$ satisfying the following properties:
\begin{gather}
  \new a{:}\vty.\forall  N{:}\ety.~{var(a)}\subs{a}{N} \eq N\\
  \new a{:}\vty,b{:}\vty.\forall  N{:}\ety.~ {var(b)}\subs{a}{N} \eq var(b) \\
  \new a{:}\vty.\forall N, M_1,M_2{:}\ety.~ {app(M_1,M_2)}\subs{a}{N} \eq app({M_1}\subs{a}{N}, {M_2}\subs{a}{N})\\
  \new a{:}\vty.\new b{:}\vty.\forall N,
  M{:}\ety.~b \fresh N \impp {lam(\abs{b}{M})}\subs{a}{N} \eq
  lam(\abs{b}{{M}\subs{a}{N}})
\end{gather}

The $\beta$-reduction and $\eta$-reduction relations are also
definable, using the following formulas:
\[\begin{array}{lc}
  (\beta) & app(lam(\abs{a}{M}),N) \to_{\beta} M\subs{a}{N}\\
  (\eta)&  \forall M.
\new a. lam(\abs{a}{app(M,a)}) \to_\eta M
\end{array}\] 
along with reflexivity, transitivity, symmetry, and congruence
properties, if desired.  Note that the implicit constraint $a \fresh
M$ arising from the quantifier ordering in $(\eta)$ corresponds to
the traditional side-condition $a \not\in FV(M)$ on $\eta$-reduction.

We now prove two elementary properties of capture-avoiding
substitution.

\begin{proposition}\labelProp{fresh-substitution}
$\models \forall a{:}\vty,N,M{:}\ety.~a \fresh M \impp {M}\subs{a}{N} \eq M$.
\end{proposition}
\begin{proof}
  Proof is by the structural induction principle $(\Lambda_{ind})$
  applied to $M$.  If $M\eq var(b)$, then we must have $a \fresh b$
  since $var$ is injective.  So ${M}\subs{a}{N} \eq{var(b)}\subs{a}{N}
  \eq var(b) \eq M$.  If $M \eq app(M_1,M_2)$, then $a \fresh M$
  implies $a \fresh M_1,M_2$, so by induction,
  ${app(M_1,M_2)}\subs{a}{N} \eq
  app({M_1}\subs{a}{N},{M_2}\subs{a}{N}) \eq app(M_1,M_2)$.  If $M \eq
  \abs{\Ab}{M'}$ for some fresh $\Ab \fresh a,N$, then $a \fresh M'$, so
  by induction, ${M}\subs{a}{N} \eq {lam(\abs{\Ab}{M'})}\subs{a}{N} \eq
  lam(\abs{\Ab}{{M'}\subs{a}{N}}) \eq lam(\abs{\Ab}{M'}) \eq M$.
\end{proof}

\begin{proposition}
  $ \models \new a,b{:}\vty.\forall M,N,N'{:}\ety.~a \fresh N' \impp {M}\subs{a}{N}\subs{b}{N'} \eq
  {M}\subs{b}{N'}\subs{a}{{N}\subs{b}{N'}}\;.$
\end{proposition}
\begin{proof}
  Let $\Aa,\Ab$ be fresh names.  Proof is by the structural induction
  principle $(\Lambda_{ind})$ applied to $M$.  If $M \eq var(c)$, then
  there are three cases, depending on whether $c \eq \Aa$, $c \eq
  \Ab$, or $\Aa \fresh c \fresh \Ab$.  If $c \eq \Aa$, then ${M}\subs{\Aa}{N}
  \eq N$ and ${M}\subs{\Ab}{N'}\eq var(\Aa)$, so
\begin{eqnarray*}
{{M}\subs{\Aa}{N}}\subs{\Ab}{N'} 
\eq {N}\subs{\Ab}{N'}
\eq {var(\Aa)}\subs{\Aa}{{N}\subs{\Ab}{N'}} 
\eq {{M}\subs{\Ab}{N'}}\subs{\Aa}{{N}\subs{\Ab}{N'}}
\end{eqnarray*}
If $c \eq \Ab$, then 
\begin{eqnarray*}
{{M}\subs{\Aa}{N}}\subs{\Ab}{N'} 
\eq {var(\Ab)}\subs{\Ab}{N'} 
\eq N' 
\eq {N'}\subs{\Aa}{{N}\subs{\Ab}{N'}} 
\eq {{M}\subs{\Ab}{N'}}\subs{\Aa}{{N}\subs{\Ab}{N'}}
\end{eqnarray*}

where $N' = {N'}\subs{\Aa}{{N}\subs{\Ab}{N'}}$ because $\Aa \fresh N'$ (by
\refProp{fresh-substitution}).  If $\Aa \fresh c \fresh \Ab$ then
\begin{eqnarray*}
{{var(c)}\subs{\Aa}{N}}\subs{\Ab}{N'} 
\eq var(c) 
\eq {{var(c)}\subs{\Ab}{N'}}\subs{\Aa}{{N}\subs{\Ab}{N'}}
\end{eqnarray*}

Next, if $M = app(M_1,M_2)$, then by induction we have
\begin{eqnarray*}
{{M_1}\subs{\Aa}{N}}\subs{\Ab}{N'} &\eq&
  {{M_1}\subs{\Ab}{N'}}\subs{\Aa}{{N}\subs{\Ab}{N'}}\\
{{M_2}\subs{\Aa}{N}}\subs{\Ab}{N'} &\eq&
  {{M_2}\subs{\Ab}{N'}}\subs{\Aa}{{N}\subs{\Ab}{N'}}
\end{eqnarray*}
so we can calculate that 
\begin{eqnarray*}
{{app(M_1,M_2)}\subs{\Aa}{N}}\subs{\Ab}{N'} 
&\eq& app({{M_1}\subs{\Aa}{N}}\subs{\Ab}{N'},{{M_2}\subs{\Aa}{N}}\subs{\Ab}{N'})\\
&\eq& app({{M_1}\subs{\Ab}{N'}}\subs{\Aa}{{N}\subs{\Ab}{N'}},{{M_2}\subs{\Ab}{N'}}\subs{\Aa}{{N}\subs{\Ab}{N'}})\\
&\eq& {{app(M_1,M_2)}\subs{\Ab}{N'}}\subs{\Aa}{{N}\subs{\Ab}{N'}}
\end{eqnarray*}

Finally, for the case of $\lambda$-abstraction, suppose that $\Ac
\fresh \Aa,\Ab,N,N'$ and $M \eq lam(\abs{\Ac}{M'})$; the induction
hypothesis is
\begin{equation*}
  {{M'}\subs{\Aa}{N}}\subs{\Ab}{N'} \eq
  {{M'}\subs{\Ab}{N'}}\subs{\Aa}{{N}\subs{\Ab}{N'}}\;.
\end{equation*}
Under these assumptions, we can conclude
\begin{eqnarray*}
{{lam(\abs{\Ac}{M'})}\subs{\Aa}{N}}\subs{\Ab}{N'} 
&\eq& lam(\abs{\Ac}{{{M'}\subs{\Aa}{N}}\subs{\Ab}{N'}})\\
&\eq&   lam(\abs{\Ac}{{{M'}\subs{\Ab}{N'}}\subs{\Aa}{{N}\subs{\Ab}{N'}}})\\
&\eq&   {{lam(\abs{\Ac}{M'})}\subs{\Ab}{N'}}\subs{\Aa}{{N}\subs{\Ab}{N'}}\;.
\end{eqnarray*}
Note that the last step relies on the fact that since $\Ac \fresh
N,\Ab,N'$, it follows that $\Ac \fresh N\subs{\Ab}{N'}$.  This
completes the proof.
\end{proof}

The above proof should seem trivial, and this is the point: nominal
abstract syntax facilitates a rigorous style of reasoning with names
and binding that is close to intuition and informal practice.
Moreover, it provides an equational theory for dealing with
expressions involving names and binding using standard algebraic and
logical techniques.  This advantage is shared by name-free approaches
such as combinatory logic or de Bruijn indices.  However, the latter
approaches rely on cleverly getting rid of explicit names.  As a
result, it can be awkward or impossible to reason about situations
involving free names, and even when possible, such reasoning is very
unlike informal reasoning.  In contrast, we can reason directly with
free names in nominal abstract syntax in a formal, yet intuitive way.

More advanced approaches such as higher-order abstract syntax often
provide properties like the substitution lemma above ``for free'',
that is, as a consequence of the metatheory of the higher-order
metalanguage.  This means that for languages whose metatheory does not
match that of the metalanguage, such properties must again be proved
in detail, and higher-order abstract syntax is too high-level for
this, since names and fresh name generation are no longer accessible.
In contrast, nominal techniques require explicit definitions of
and reasoning about substitution, but are more flexible.

\section{Applications}

\subsection{Programming techniques}

Probably the most immediately useful application of nominal abstract
syntax is in providing more advanced support for symbolic programming
with languages with bound names.  One obvious approach is to extend an
existing programming language. with support for nominal abstract
syntax, much as the languages \lprolog, Twelf,
$ML_\lambda$~\cite{miller90lf}, and Delphin~\cite{schurmann04unpub}
extend logic or functional programming paradigms with support for
higher-order abstract syntax.

FreshML~\cite{pitts00mpc,shinwell03icfp,shinwell03merlin} is an
extension of the ML programming language that provides built-in
support for nominal abstract syntax. The main additions to the
language are the \texttt{let x:name = fresh} construct, which chooses
a fresh name and binds it to $x$, and the abstraction type/term
constructor.  Abstractions are considered equal up to
$\alpha$-renaming, and pattern matching against abstractions
automatically freshens the bound name.  In early versions of FreshML,
a complicated type analysis was employed to ensure that
name-generating functions were ``pure'' (side-effect-free); this
analysis was found to be overly restrictive and has been dropped in
recent versions resulting in a language that is more permissive but
has side-effects.  There are no constant names in FreshML; instead,
names are always obtained via fresh name generation and manipulated
via variables.  On the other hand, in recent versions of FreshML,
names may be attached to data such as strings and integers; also, data
structures containing names may be bound, not just individual names.
The current implementation is available as an extension to the
Objective Caml language, and so inherits the mature compiler and
libraries available for that language.

\refFig{freshml} shows an interesting example FreshML program (from
Shinwell et al.~\cite{shinwell03icfp}).  It implements
\emph{normalization by evaluation}, an advanced technique for
optimization that uses ML's higher-order features to simplify
lambda-calculus expressions.  In ordinary ML, fresh names must be generated by a
user-defined $gensym$ function; in contrast, in FreshML, built-in
binding and fresh name generation can be used.  In addition, although
this program uses side-effects, it is not difficult to prove that it
is actually a pure function (as one would expect).
\begin{figure}[tb]
\begin{verbatim}
datatype lam = Var of name 
             | Lam of <name>lam
             | App of lam * lam;

datatype sem = L of (unit -> sem) -> sem
             | N of neu
and      neu = V of name
             | A of neu * sem;

fun reify(L f)     = let fresh x:name in 
                     Lam(<x>(reify(f(fn () => N(V x)))))
  | reify(N n)     = reifyn n
and reifyn(V x)    = Var x
  | reifyn(A(n,d)) = App(reifyn n, reify d);

fun evals [] (Var x)          = N(V x)
  | evals((x,v)::env) (Var y) = if x = y then v()
                                else evals env (Var y)
  | evals env (Lam(<x>t))     = L(fn v => evals ((x,v)::env) t)
  | evals env (App(t1,t2))    = case evals env t1 
                                  of L f => f(fn () => evals env t2)
                                   | N n = N(A(n, evals env t2));

fun eval t = evals [] t;
fun norm t = reify(eval t);
\end{verbatim}
\caption{Normalization by evaluation in FreshML}\labelFig{freshml}
\end{figure}

\aprolog is a Prolog-like language that supports nominal abstract
syntax; roughly speaking, it is to Prolog as FreshML is to ML.  Unlike
ordinary Prolog, \aprolog is strongly typed.  Some simple Horn clause
programs involving $\lambda$-terms were shown in \refFig{aprolog}.
Types in \aprolog are useful for describing the binding structure of
term languages, and help catch many more errors statically.  The
unification algorithm used in \aprolog is essentially the
\emph{nominal unification} algorithm developed by Urban, Pitts, and
Gabbay~\cite{urban04tcs}.  In this algorithm (and in \aprolog) names
are treated as concrete constants, rather than requiring that names
are only manipulated via variables.  However, name constants in
program clauses are not interpreted as global constants, but as
$\new$-quantified within the clause.  Thus, two names appearing in the
clause may be instantiated with any two other distinct names, but not
with the same name.  In particular, this means that freshness and
inequality constraints between names and other data can be employed to
correctly implement informal freshness side-conditions, as shown in
the closure conversion example of \refFig{aprolog-cc}.

\begin{figure}[tb]
\[\begin{array}{lll}
  cconv([\Ax|G],var(\Ax),Env,pi_1(E)).\\
  cconv([\Ax|G],var(\Ay),Env,E') &\ent& cconv(G,var(\Ay),pi_2(Env),E').\\
  cconv(G,app(T_1,T_2),Env,E')) &\ent& cconv(G,T_1,E_1),cconv(G,T_2,Env,E_2),\\
&&E' = let(E_1,\abs{\Ac}{ app(pi_1(var(\Ac)),pair(E_2,pi_2(var(\Ac))))}. \\

\multicolumn{3}{l}{cconv(G,lam(\abs{\Ax}T),Env,pair(lam(\abs{\Ay}E'),E))}\\
 &\ent& \Ax\fresh G,\Ay\fresh G,\\
&&cconv([\Ax|G],T,var(\Ay),E').
\end{array}\]%
\caption{Closure conversion in \aprolog}\labelFig{aprolog-cc}
\end{figure}

\aprolog is an example of a \emph{nominal logic programming language},
that is, its logical foundation is nominal logic.  Initially, the
connection between nominal logic as originally formulated by Pitts and
the operational behavior of unification and proof search in \aprolog
was less than clear.  Nominal logic as reformulated by
Cheney~\cite{cheney04phd,cheney05jsl} (and presented in simplified
form in this paper) now provides a robust logical and semantic
foundation for \aprolog; however, as currently implemented, \aprolog
is logically incomplete (that is, there are queries whose answers
cannot be found, even in principle.)  The reason is that \aprolog's
proof search and unification techniques address only the equational
theory of nominal logic.  Because of the equivariance principle, this
is not enough; for example, the two atomic formulas $p(\Aa)$ and
$p(\Ab)$ are \emph{logically equivalent} but not provably equal as terms.

There are two ways around this: solve the more general problem, or
limit the programs so that the special case is enough.  Both
approaches have been explored.  Complete proof search for general
nominal logic programs requires solving \emph{equivariant unification}
problems~\cite{cheney04icalp,cheney05rta}, that is, unifying atomic
formulas up to both a substitution for free variables and a
name-permutation.  This process is $\NPTIME$-complete and appears
nontrivial to implement.  Urban and Cheney~\cite{urban05tlca} studied
a fragment of nominal logic programs for which proof search based on
nominal unification \emph{is} complete.  The idea of this result is
that clauses that only manipulate bound names in simple ways are
\emph{automatically} equivariant, so explicit unification modulo
equivariance is unnecessary.  Although it excludes some interesting
programs (such as closure conversion), this fragment includes many
interesting \aprolog programs, though slight modifications are
sometimes necessary.

It is important to point out that higher-order abstract syntax offers
benefits for programming with names and binding that nominal abstract
syntax still lacks; in particular, HOAS provides capture-avoiding
substitution as an efficient, built-in operation, whereas nominal techniques
typically do not.  But higher-order techniques seem much more
difficult to incorporate into existing languages, because of the need
for higher-order unification and matching to manipulate higher-order
terms.  On the other hand, while capture-avoiding substitution is
annoying to have to implement in FreshML or \aprolog, there is no
conceptual problem in doing so; instead, the problem is that there are
a large number of similar ``boilerplate'' cases that are conceptually
uninteresting but have to be written anyway.

The need to write such boilerplate code and the need to switch to a
new language are two potential drawbacks for programmers considering
nominal abstract syntax.  Recently, some progress has been made on
both fronts, by projects providing nominal abstract syntax as a
library or lightweight language translation rather than as a language
extension, and employing \emph{generic programming} techniques to
alleviate the burden of implementing name-related boilerplate (or
\emph{nameplate}).  Cheney~\cite{cheney05icfp} developed a Haskell
class library called \emph{FreshLib} that provides much of the
functionality of FreshML within Haskell, and also showed how to use
L\"ammel and Peyton~Jones' \emph{scrap your boilerplate} approach to
generic programming~\cite{lammel05icfp} to provide reusable, generic
definitions of capture-avoiding substitution and other nameplate.
Pottier~\cite{pottier05ml} has developed C$\alpha$ml, a language tool
for Objective Caml that translates type declarations decorated with
binding specifications to plain Objective Caml programs that
automatically deal with name-binding.  C$\alpha$ml's binding
specifications can describe more complex binding structure than the
one-name-at-a-time binding present in nominal logic; for example,
C$\alpha$ml can express pattern matching and $letrec$ binding forms.
In addition, C$\alpha$ml provides \emph{visitor} classes that can
easily be overridden to implement capture-avoiding substitution.
While these developments are encouraging, it is not clear yet whether
nominal techniques can provide the same combination of convenience and
efficiency as higher-order techniques.

\subsection{Automated reasoning}

A second major application area for nominal techniques is specifying
and proving properties of formal systems, including logics,
programming language calculi, concurrency calculi, and security
protocols.

Initial work of this form was carried out by Gabbay.
Gabbay~\cite{gabbay01phd} implemented FM set theory in the Isabelle
theorem prover, as a variant of Isabelle's implementation of ZF set
theory (Isabelle/ZF).  Unfortunately, Isabelle/ZF relies heavily on
the Axiom of Choice, even in places where it is not strictly
necessary.  Because FM set theory is incompatible with the Axiom of
Choice, it was necessary to re-develop a significant amount of set
theory in Isabelle/FM.  Gabbay also investigated FM-HOL, a form of
higher-order logic based on FM set theory~\cite{gabbay02automath}.  As
far as I know this has not been implemented; implementing FM-HOL as a
variation on Isabelle/HOL would likely involve considerable effort
because Isabelle/HOL also relies extensively on the Axiom of Choice.

Gabbay's work was \emph{foundational} in the sense that it attempted
to incorporate nominal techniques into the surrounding mathematical
foundations.  While this is attractive because it builds several
desirable properties into the foundations, it has a high start-up cost
and requires potential users to adapt to the new foundations.  One
motivation for Pitts' development of nominal logic was to show how to
work with nominal abstract syntax without leaving the mathematical
mainstream.  Pitts' recent paper~\cite{pitts05tphols} continues this
theme by showing how to relate classical and nominal approaches to
$\alpha$-equivalence.  Based on this insight,
Norrish~\cite{norrish04tphol}, Urban and Tasson \cite{urban05cade} and
Urban and Norrish~\cite{urban05merlin} have performed
\emph{non-foundational} formalizations of properties of the
$\lambda$-calculus in classical higher-order logic. Rather than build
swapping, freshness, etc. into the logic, Urban and others have
defined swapping functions and constructed nominal abstract syntax
trees explicitly, and proved explicit induction or recursion
principles.  Although some additional subgoals need to be proved in
this approach relative to a foundational approach, the start-up cost
of implementing this approach appears lower, and the learning curve
for users already familiar with Isabelle/HOL seems gentler.  Urban and
others are currently working on extending Isabelle/HOL's datatype
package so that induction and recursion principles can be derived
automatically for datatypes employing nominal abstract syntax.

Nominal logic appears to be related to two other recently investigated
approaches to formal reasoning about languages with names: the Theory
of Contexts~\cite{honsell01icalp} and \FOLNabla~\cite{miller03lics}.  Both
have been used to carry out complete machine-checkable formalizations
of properties of interesting languages, including the $\pi$-calculus.

The Theory of Contexts (TOC) is an axiomatic extension to the Calculus
of Inductive Constructions (or CIC), the type theory of the Coq
system.  In TOC, names are represented by an abstract base type $V$.
Equality is assumed to be decidable for names.  Moreover, there is a
freshness relation $notin$ relating names and arbitrary values, and
fresh names are always assumed to exist.  Name-binding is represented
using the function type $V \to X$.  The Theory of Contexts appears to
be similar in some respects to nominal logic; in fact, Miculan,
Scagnetto, and Honsell~\cite{miculan05merlin} have developed a
translation from nominal logic specifications to TOC specifications.
This translation is sound (translates derivable formulas of nominal
logic to derivable formulas of CIC/TOC) but not complete (some
non-derivable formulas have derivable translations).  This is because
the Theory of Contexts is set in a higher-order type theory that is
stronger than the first-order setting of nominal logic.  Nevertheless,
it seems fair to say that TOC is approximately equivalent to nominal
logic in expressive power.  The chief difference seems to be the
handling of binding via a higher-order function encoding rather than
an explicit axiomatization.

Miller and Tiu~\cite{miller03lics} have introduced \FOLDNabla, which
stands for \underline{F}irst-\underline{O}rder logic with
\underline{$\lambda$}-terms, \underline{D}efinitions, and the
\underline{$\nabla$}-quantifier.  As its title suggests, this logic
includes function types populated by $\lambda$-terms, but only permits
quantification over ``first-order'' types (that is, types not
mentioning $o$, the type of propositions).  In addition, \FOLDNabla
includes the ability to make definitions and perform case-based
reasoning on the structure of definitions, and a novel self-dual
quantifier $\nabla$.  Though $\nabla$ and $\new$ behave differently,
Gabbay and Cheney~\cite{gabbay04lics} developed a partial (sound but
incomplete) translation from \FOLNabla (the definition-free part of
\FOLDNabla) to nominal logic, and Cheney~\cite{cheney05fossacs} has
developed an improved, sound and complete translation.
Miculan and Yemane~\cite{miculan05fossacs} have investigated the idea
of using the semantics of nominal logic as the basis of a denotational
semantics for \FOLNabla.

Computational techniques such as unification, constraint solving, and
rewriting are very relevant to automated deduction.  As noted earlier,
Urban, Pitts, and Gabbay~\cite{urban04tcs} first studied unification
for nominal terms.
Cheney~\cite{cheney04icalp,cheney04phd,cheney05rta} noted that the
unification problems their algorithm solves are only special cases of
the problems that must be solved in general nominal logic programming
or rewriting.  The general cases of nominal unification
(solving equations $t \eq u$), freshness constraint solving ($a \fresh
t$), and equivariant unification ($\exists \pi. \pi \act t \eq u$) are
all $\NPTIME$-complete.  In fact, even equivariant matching (that is,
equivariant unification where one side is ground) is
$\NPTIME$-complete.  Despite this, several tractable special cases of
these problems are known.

A significant area for future work in this area is the investigation
of \emph{nominal equational unification}: that is, unification modulo
an extension to the equational theory of nominal logic.  Many
structural congruences considered in concurrency calculi 
can be expressed by a nominal equational theory.  For
example, $\pi$-calculus terms are considered structurally congruent
modulo axioms such as:
\[x \fresh P \impp nu(\abs{x}P) \eq P \qquad nu(\abs{x}{nu(\abs{y}{P})}) \eq nu(\abs{y}{nu(\abs{x}{P})}) \]
Fernandez, Gabbay, and Mackie~\cite{fernandez04ppdp} have investigated
nominal term rewriting systems.  They study conditions for
establishing confluence and show how existing higher-order rewriting
formalisms can be simulated using nominal rewriting.  They also
discuss the implications of the $\NPTIME$-hardness of equivariant
matching and present a syntactic condition on rewriting rules that
ensure nominal matching is sufficient.  Fernandez and
Gabbay~\cite{fernandez05ppdp} investigate an extension of nominal
rewriting with a ``hidden name'' operation $\new a.t$; this operation
behaves like the name-restriction operation $\nu a.P$ in the
$\pi$-calculus.  Gabbay~\cite{gabbay05ppdp} has also proposed a novel
approach to reasoning about contexts (that is, terms with ``holes'')
based on nominal techniques.  

So far most techniques for automating reasoning with nominal abstract
syntax have focused on general-purpose formal systems and proof tools
such as Isabelle/HOL or Coq.  In contrast, there are several
lightweight, domain-specific applications for formalizing, programming
with, and reasoning about various forms of higher-order abstract
syntax, including \lprolog~\cite{nadathur98higher},
Twelf~\cite{pfenning99system}, Linear LF~\cite{cervesato02ic}, and
Concurrent LF~\cite{watkins03types}.  These systems seem to have a
much gentler learning curve than the general-purpose systems, yet are
suitable for a wide range of applications.  I am particularly
interested in developing an analogous \emph{logical framework} for
nominal abstract syntax.  The \aprolog language can be viewed as a
first step in that direction.

On the other hand, the fact that nominal abstract syntax can be
constructed explicitly in general-purpose reasoning systems such as
Isabelle/HOL opens up another new direction: relating denotational and
operational semantics, and proving results via denotational rather
than via operational techniques.

\subsection{Semantics}

Besides providing a foundation for programming and reasoning in the
presence of name-binding, nominal techniques have several applications
in logic and programming language semantics.  

Early work by Pitts and
Stark~\cite{pitts93mfcs} and Odersky~\cite{odersky94popl} studied
name-generation in functional programming languages as a simplified
case of general side-effects.
Pitts and Stark's nu-calculus~\cite{pitts93mfcs} analyzed name
generation as an effectful computation, analogous to reference
generation in ML.  Names could be introduced with a ``fresh name''
binder $\nu n.t$, and tested for equality.  The fresh name
construction was interpreted operationally by maintaining a
name-store: on encountering a $\nu n.t$ term, a fresh name is bound to
$n$ and added to the store.  In some ways, this work can be seen as an
early precursor to that of Pitts and Gabbay on
FreshML~\cite{pitts00mpc}.
Odersky developed a quite different functional theory of local
names~\cite{odersky94popl}.  His $\lambda\nu$-calculus is
syntactically essentially the same as Pitts and Stark's nu-calculus.
But instead of treating name-generation as an effect, the
$\lambda\nu$-calculus deals with names in a local and functionally
pure way.  Odersky developed a denotational
semantics for $\lambda\nu$ in terms of name-swapping and support.
This development clearly foreshadows the later developments underlying
nominal logic, FreshML, and \aprolog, although, of course, without the
application to binding.

Sch\"opp and Stark~\cite{schopp04csl} have developed a form of type
theory based on nominal logic and the logic of bunched implications
(BI)~\cite{pym02semantics}. Intuitively, the idea of this system is to
identify the ``resources'' of BI with the sets of names supporting
values in nominal logic.  This theory axiomatizes a type of names and
includes ``fresh'' dependent products and sums $\Pi^*, \Sigma^*$
(corresponding to $\forall^{new}, \exists^{new}$ in BI) in addition to
ordinary dependent products and sums.  Though very interesting, there
are many unresolved practical problems in this setting (for example,
it is not yet known whether strong normalization holds).

One area in which reasoning about name-generation is of particular
interest is in concurrency calculi, in particular the $\pi$-calculus.
A number of researchers have investigated applications of nominal
techniques to the $\pi$-calculus and similar
systems~\cite{gabbay03automath,ferrari05fossacs,stark05fossacs}.  Ideas from
nominal logic have also been incorporated into logics for reasoning
about concurrency or data
structures with name-hiding~\cite{caires02concur,cardelli03fossacs}.

\section{Future directions}

The equivariance principle is an integral component of nominal logic
as currently formulated.  Without it, the $\new$-quantifier would no
longer be self-dual, and reasoning about function and relation symbols
would be significantly more complicated.  On the other hand, it has
several undesirable consequences.  As noted earlier, because of
equivariance, \aprolog's intuitively appealing proof search strategy
based on nominal unification is incomplete; to obtain completeness, it
is necessary to either place significant restrictions on the language
or solve $\NPTIME$-complete equivariant unification problems.
Furthermore, equivariance implies that no linear order on the set of
names (or even any finite nonempty subset) can be denoted by a
relation symbol, since $a < b$ and $b < a$ cannot both hold.  For this
reason, I believe that finding an alternative approach to nominal
logic that supports names, $\new$-quantification, and binding without
relying on equivariance is an important open issue.

Model theory and database theory~\cite{libkin04finite} may be
interesting places to look for inspiration concerning how to attack
this problem.  In model theory, the groups of automorphisms or
order-preserving automorphisms play important roles.  In database
theory, \emph{generic} queries whose answers are invariant under
permutations of the domain elements are often of interest.  Thus,
generic queries are similar to equivariant formulas.  In addition, a
great deal of research has concerned the impact of having a total
order on the domain of individuals on expressive power.  In this
setting, one typically considers \emph{order-invariant} queries whose
meaning is independent of the linear order.  

Another important direction for future work is to reconcile nominal
logic with well-known constructive or type-theoretic principles.  I
believe nominal abstract syntax is entirely satisfactory from the
point of view of mathematical constructiveness; for example, nominal
abstract syntax trees can be defined as an inductive construction
similar to that used for ordinary abstract syntax trees, and by design
nominal logic does not rely on the Axiom of Choice.  Moreover, the
proof theory and semantics of intuitionistic nominal logic has been
investigated by Gabbay and
Cheney~\cite{gabbay03unpub,gabbay04lics,cheney05fossacs}.  However,
nominal logic seems challenging to integrate with type-theoretic
approaches to computation and reasoning, because of its use of
explicit freshness constraints, its non-confluent equational theory,
and explicit fresh name generation.  Some of these problems have
already been encountered in purely functional versions of FreshML and
in Sch\"opp and Stark's dependent type theory with names.

\section{Conclusion}

Gabbay and Pitts' approach to formalizing abstract syntax with names
and binding (i.e., \emph{nominal abstract syntax}) is an important
foundational development relevant to logic and computer science.  It
provides a level of abstraction for reasoning about languages with
binding that lies between first-order abstract syntax (which is
usually too low-level) and higher-order abstract syntax (which is
powerful, but too high-level for some applications).  Nominal abstract
syntax also seems to provide justification for the kind of reasoning
people already perform, rather than requiring new proof techniques.

Nominal logic is an extension of first-order logic formalizing the
principles of nominal abstract syntax.  It has numerous applications,
ranging from providing a foundation for a logic programming language
to machine-checked or assisted proofs of language properties.  In this
column I have attempted to give an impression of the key ideas
underlying nominal logic, how it can be used, how it is being
applied, and how it might be improved.

\bibliographystyle{plain}

\end{document}